\documentclass[%
 reprint,
 amsmath,amssymb,
 aps,
]{revtex4-2}

\usepackage{graphicx}
\usepackage{dcolumn}
\usepackage{bm}

\begin{document}

\preprint{APS/123-QED}

\title{Shape Elasticity in Colloidal Bent-Core Liquid Crystals}

\author{Nicholas W. Hackney}
\email{nwhackn@sandia.gov}
\author{Joel T. Clemmer}
\author{Gary S. Grest}
\affiliation{Sandia National Laboratories, Albuquerque, New Mexico 87185, USA}%

\date{\today}

\begin{abstract}
Curved particles have been shown to stabilize a range of states with unique order in dense suspensions of colloidal bent core liquid crystals. The shape of the colloidal rods encourages the formation of curved director fields. However, states of constant bend cannot uniformly fill either two or three dimensional Euclidean space and are therefore geometrically frustrated. As a result, curved rods are forced to couple their preference for bend with additional twist and splay deformations, giving rise to twist-bend and splay-bend states of nematic and smectic order. In this article, we study the effect of rod curvature on these diverse states of
liquid crystalline order using molecular dynamics simulations of a bonded particle model of curved rods with tunable shape elasticity. Focusing on the case of intermediately curved rods, we find that curved rods go through a sequence of isotropic, nematic twist-bend and smectic splay-bend ordering as the density is increased from the dilute limit, in agreement with previous studies of rigid rods. As the rods become more elastic, the critical concentration separating these phases is shifted to higher density. Lastly, we find that flexibility weakens the first-order phase transition separating the isotropic and nematic twist-bend phases.
\end{abstract}

\maketitle

\section{Introduction}

Colloidal liquid crystals consist of suspensions of elongated rod-like particles in a solvent. In contrast to thermotropic liquid crystals\textemdash whose phase behavior is controlled by temperature, these colloidal systems are lyotropic, which means that their phase behavior is largely controlled by the concentration of rod-like particles.~\cite{fernandez2024liquid} Such colloidal liquid crystals have been studied in the context of suspensions of rod-like viruses,\cite{tang1995isotropic,dogic2000cholesteric,dogic2001development} SU-8 epoxy~\cite{fernandez2019synthesis} and silica~\cite{kuijk2011synthesis,wittmann2021particle} rods, natural clay rods,\cite{zhang2006isotropic} cellulose nano-crystals,\cite{revol1992helicoidal,parton2022chiral} and DNA origami tubules.\cite{siavashpouri2017molecular} In these systems, it is well known that shape anisotropy leads to the formation of a variety of states with nematic, smectic and chiral liquid crystalline order~\cite{fernandez2024liquid}. This diverse phase behavior has made colloidal liquid crystals excellent model systems for designing complex functional material with a range of novel optical~\cite{schutz2020equilibrium,dumanli2014controlled} and topological~\cite{fernandez2021hierarchical,khanra2022controlling} properties. Recent advances in fabrication techniques~\cite{fernandez2020shaping,yang2016synthesis,wu2017colloidal} have allowed the study of curved or bent-core liquid crystals. 

Curvature breaks the local mirror symmetry of the individual rods.\cite{dozov2001spontaneous,xu2020general} As a result, these systems require three orthonormal directors to describe the order: a nematic director ($\mathbf{n}$), a bend director ($\mathbf{b}$) and an out-of-plane director $\mathbf{m}=\mathbf{n}\times\mathbf{b}$.  The curvature of the rods encourages them to form arrangements with uniformly curved nematic director field at high density. However, it has been shown that such curved arrangements cannot smoothly tile Euclidean space and these bent-core liquid crystals are therefore geometrically frustrated. Thus, local preference of the rods to bend must be accompanied by additional splay and twist deformation modes,\cite{meyer1976structural,edwards1976molecular,dozov2001spontaneous} giving rise to a variety of ordered phases. Additionally, coupling between twist and bend has been theorized~\cite{binysh2020geometry} to allow for the formation of defective double twist cylinders (i.e. Skyrmion or meron-like structures) that have been shown to arrange into a variety of ordered and disordered blue phases ($BP$) in cholesteric liquid crystals.\cite{PhysRevX.12.011003,wright1989crystalline}

The effect of this frustration on the equilibrium behavior of thermotropic bent-core liquid crystals has been the subject of many theoretical,\cite{lubensky2002theory} computational,\cite{shamid2013statistical,shamid2014predicting} and experimental~\cite{takezoe2006bent,jakli2018physics,jakli2023defects} studies. However, it has not been until recently that the role of concentration on the phase behavior of colloidal bent-core liquid crystals has begun to be understood. Notably, there have been several studies that have used a combination of theoretical~\cite{anzivino2020landau,anzivino2022coupling} and computational~\cite{memmer2002liquid,greco2015entropy,kubala2022silico,chiappini2021generalized} techniques to investigate the concentration-curvature plane of phase space of bent-core rods with uniform opening angle, $\Psi=L/R$. Here, $L$ and $R$ correspond to the length and radius of curvature of the rods, respectively. Through tuning the curvature and density, they showed the existence of isotropic ($I$); nematic ($N$); nematic twist-bend ($N_{\rm TB}$); and smectic-A ($Sm_{\rm A}$), splay-bend ($Sm_{\rm SB}$) and polar ($Sm_{\rm X}$) states of order. The arrangement of these phases under increasing concentration can be summarized as follows: in the limit of low curvature, the rods behave as if they are effectively straight and go through a sequence of $I\rightarrow N\rightarrow Sm_{\rm A}$ order. As the curvature increases, the $N\rightarrow Sm_{\rm A}$ transition is interrupted by the appearance of a $N_{\rm TB}$ phase, which transitions to a $Sm_{\rm SB}$ phase at high concentration. At intermediate curvatures, the rods never behave as if they are straight and go through the sequence $I\rightarrow N_{\rm TB}\rightarrow Sm_{\rm SB/X}$ as the concentration increases. Molecular dynamics (MD) simulations investigating the intermediate curvature regime observe the formation of double twist cylinders, which can form stable hexagonal lattices in quasi-two-dimensional confinement and disordered blue phases that are at-least metastable in bulk.\cite{subert2024achiral} Lastly, in the very high curvature regime, the rods begin to close into annular rings which behave as colloidal disks ordering into columnar stacks.\cite{wu2017colloidal} To date, many of these smectic~\cite{yang2016synthesis,yang2018phase,fernandez2020shaping} and columnar~\cite{wu2017colloidal} phases have been observed experimentally. However, the nematic twist-bend phase has yet to be observed in colloidal liquid crystals and remains an open challenge.\cite{fernandez2024liquid}

Previous computational studies of colloidal bent-core liquid crystals considered only rigid rods and ignored the effects of elastic deformation of their shape. As a result, the effect that elasticity has on the equilibrium phase behavior is not well understood. However, rod flexibility could have several important effects, altering the energetic competition between the bend, twist, and splay deformation modes. This could effectively change the conditions under which each phase is stable, making these phases more or less accessible. In addition, variation of the competition between different deformation modes could stabilize additional structures not seen in the equilibrium phase behavior of rigid rods. For example, the heliconical twist-bend phase and the double twist, Skyrmion-like defects both arise from a coupling of twist and bend. However, the heliconical state has a constant magnitude of bend, whereas double twist cylinders have coherent bend gradients~\cite{binysh2020geometry}, with the magnitude decaying radially from the central axis. These gradients are only possible in colloidal suspensions of rigid rods via excluded volume effects, which create an effective bending by varying the space between rods. In this case, shape elasticity could ostensibly alter the competition between twisting and bending, making double twist cylinders energetically favorable. This could, in turn, stabilize either body centered or simple cubic defect networks (i.e. the so-called BPI \& II phases) that have been found in thermotropic cholesteric liquid crystals~\cite{wright1989crystalline,tanaka2015double,nych2017spontaneous} but have yet to be seen in colloidal bent-core systems. Alternatively, rod flexibility could have a similar effect as shape dispersity, which has been shown to stabilize a window of nematic splay-bend order~\cite{chiappini2019biaxial,fernandez2020shaping,kotni2022splay} and lead to the formation of topological defect lattices in two-dimensions.\cite{fernandez2021hierarchical} Lastly, it is possible that shape elasticity could completely relieve frustration by allowing the rods to straighten.

Throughout this work, we will employ a bonded particle model of curved rods, which allows the elastic moduli of the individual colloidal particles to be tuned. Using this model, MD simulations are performed to investigate the effects of shape elasticity on the equilibrium phase behavior of colloidal bent-core liquid crystals. The rest of the article will be organized as follows. In Section \ref{sec: simulation details} we will introduce the bonded particle model and discuss its application to curved colloidal rods. In Section \ref{sec: results}, we present simulation results for a range of density and elastic constants. The liquid crystalline order is analyzed and the results are compiled into a phase diagram illustrating the effects of rod flexibility on the equilibrium behavior. In subsection~\ref{sec: fluctuations}, we analyze the elastic curvature fluctuations of the individual rods. In subsection~\ref{sec: structure}, we investigate the effect these fluctuations have on the structure of the ordered phases. Finally, in section~\ref{sec: conclusion}, we summarize these results and discuss the implications of this first investigation into the effect that shape elasticity has on the equilibrium phase behavior of colloidal bent-core liquid crystals.

\begin{figure}[ht!]%
\includegraphics[width=0.4\textwidth]{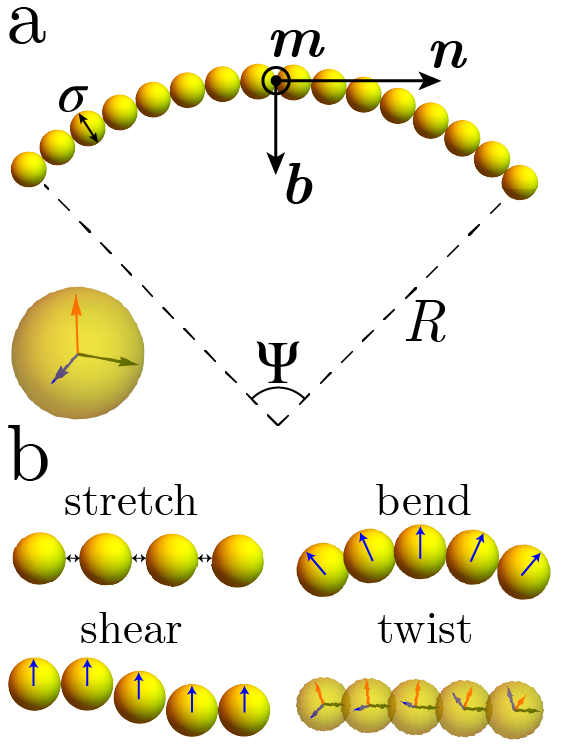}
\caption{Schematic illustration of ($\mathbf{a}$) curved rod geometry used to simulate colloidal bent-core liquid crystals. Rod geometry defined by the number of beads with diameter $\sigma$ lying along circular arc of radius $R$. The ratio of arc-length $L$ to radius of curvature defines the dimensionless opening angle, $\Psi=L/R$. The three orthonormal vectors, ${\mathbf{n},\mathbf{b},\mathbf{n}}$ needed to describe the rods orientational order are illustrated on top of the curved rod. Additionally, each bead has an internal orthonormal frame needed to describe ($\mathbf{b}$) the allowed stretching, shearing, bending and twisting modes described by the the bonded particle model. } \label{fig: BPM_Figure}
\end{figure}

\begin{figure}[ht!]%
\includegraphics[width=0.5\textwidth]{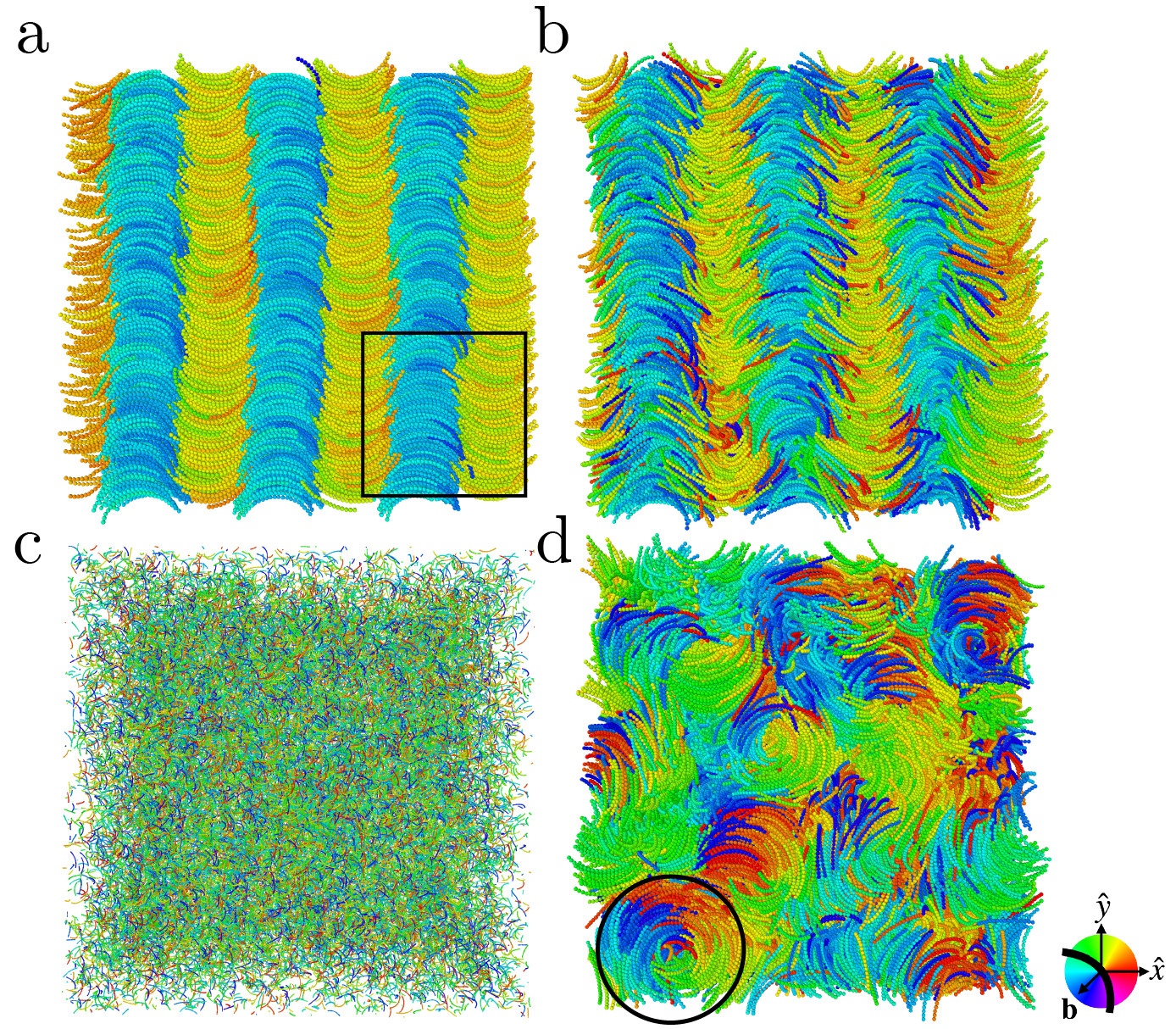}
\caption{Starting states used to initialize simulations. (\textbf{a}) An initial high density ordered phase, obtained by replicating a smaller sample (black square roughly denotes size of replicated cell), which allows the system to avoid getting stuck in meta-stable state as the pressure is increased. This state is then lowered to equilibrate to (\textbf{b}) a lower density ordered phase with $P=0.5\epsilon/\sigma^3$. Alternatively, starting from (\textbf{c}) the low density, low pressure isotropic phase and compressing to $P=0.5\epsilon/\sigma^3$ results in an intermediate (\textbf{d}) meta-stable defect state. The solid black circle highlights a Skyrmion-like double twist defect characteristic of this so-called amorphous blue phase (BPIII). These meta-stable states can be very long lived and are computationally prohibitive to anneal for even moderately large systems. Rods are colored by the direction the $xy$-projection of their bend director, $\mathbf{b}$, points to when centered on the color wheel, as illustrated in the legend. All simulations have $40500$ rods and $k_{\rm b}=1000\epsilon$. For illustrative purposes, the simulation snapshots are scaled to be equal in size and therefore are not to scale.} \label{fig: Initial States}
\end{figure}

\section{Simulation Details}\label{sec: simulation details}

Individual bent-core colloidal rods are modeled as a string of 16 spherical beads with mass $m$. The beads trace out a segment of a circular arc with length $15\sigma$ and a radius of $10\sigma$, giving the rods an opening angle of $\Psi=L/R=1.5$. Here $\sigma$ defines the unit of length. A schematic illustration of this geometry, along with the three orthogonal directors ($\mathbf{n},\mathbf{b},\mathbf{m}$) needed to describe the rod orientation, is given in Figure \ref{fig: BPM_Figure}a. This particular geometry was chosen since it is in the intermediate curvature regime and is comparable to previous numerical studies of the nematic twist-bend and smectic splay-bend phases of rigid rods.\cite{chiappini2021generalized,subert2024achiral}

To allow for elastic deformations of the rod shape, a rotational bonded particle model is used to describe the intra-rod interaction between beads.~\cite{clemmer2024soft,wang2009new} This model takes the prescribed rod geometry to be a stress-free reference state to calculate the forces between neighboring beads. Additionally, the beads have internal rotational degrees of freedom, which are illustrated in the inset of Fig. \ref{fig: BPM_Figure}a as an orthonormal framing inside a sphere, allowing each bond to transmit a normal (i.e. stretching/compression) and shear force, as well as a torsional and bending torque similar to an elastic rod~\cite{clemmer2024soft}. A schematic illustration of the deformation modes associated with each of these forces and torques is shown in Figure~\ref{fig: BPM_Figure}b. Assuming small deformations and isotropic shear, the normal and shear forces between two particles can be written as~\cite{wang2009new}
\begin{equation}
\begin{split}
f_{\rm r}&=k_{\rm r}\Delta u_{\rm r} \\
f_{\rm s}&=k_{\rm s}\Delta u_{\rm s}
\end{split}
\end{equation}
where $\Delta u_{\rm r}$ and $\Delta u_{\rm s}$ define the normal and tangential displacements relative to the reference state and $k_{\rm r}$ and $k_{\rm s}$ are the normal and shear stiffnesses, respectively and have units of force per distance (i.e. $\epsilon/\sigma^2$ in LJ units). The shear force also induces a torque between particles. Similarly, the twisting and isotropic bending torques can be written as~\cite{wang2009new}
\begin{equation}
\begin{split}
\tau_{\rm t}&=k_{\rm t}\Delta\alpha_{\rm t} \\
\tau_{\rm b}&=k_{\rm b}\Delta\alpha_{\rm b}
\end{split}
\end{equation}
where $\Delta\alpha_{\rm t}$ and $\Delta\alpha_{\rm b}$ define the angular displacement of twisting and bending relative to the reference state. Likewise, $k_{\rm t}$ and $k_{\rm b}$ are the bending and twisting stiffnesses and have units of torque per radian (i.e. $\epsilon$ in LJ units). Thus, there are four parameters, $\{k_{\rm r},k_{\rm s},k_{\rm t},k_{\rm b}\}$, controlling the elasticity of the rods within the rotational bond model.

Beads on different rods interact via a Lennard-Jones potential,
\begin{equation}
U_{\alpha\beta}=\begin{cases}
4\epsilon\big[\big(\frac{\sigma}{r}\big)^{12}-\big(\frac{\sigma}{r}\big)^{6}\big] & r<r_{\rm c} \\
0 & r>r_{\rm c}
\end{cases}
\end{equation}
where $\epsilon$ is the interaction energy. The cutoff distance ${r_{\rm c}}=2^{1/6}\sigma$ is chosen so that the rod-rod interactions are purely repulsive. Using this model, molecular dynamics simulations were carried out using the LAMMPS software package~\cite{thompson2022lammps} at constant rod number $N$, pressure $P$, and temperature $T=\epsilon/k_{\rm B}$ where $k_{\rm B}$ is the Boltzmann constant. Throughout this work, the simulation cell is orthogonal with periodic boundary conditions in all three directions. The equations of motion are integrated using the velocity-Verlet algorithm with a timestep $\delta t=0.0035\tau_{\rm LJ}$, where $\tau_{\rm LJ}=(m\sigma^2/\epsilon)^{1/2}$ is the standard time unit for a Lennard-Jones system.  A Langevin thermostat~\cite{schneider1978molecular} with damping $1\tau_{\rm LJ}$ and a Berendsen barostat~\cite{berendsen1984molecular} with damping $200\tau_{\rm LJ}$ are used to maintain the temperature and pressure, respectively. Note that the specific implementation of the Berendsen barostat used here allows for anisotropic dilation or compression of the simulation box to avoid any potential effects arising from incommensurability between box dimension and periodicity of any of the ordered phases. 

\begin{figure*}[ht!]%
\includegraphics[width=\textwidth]{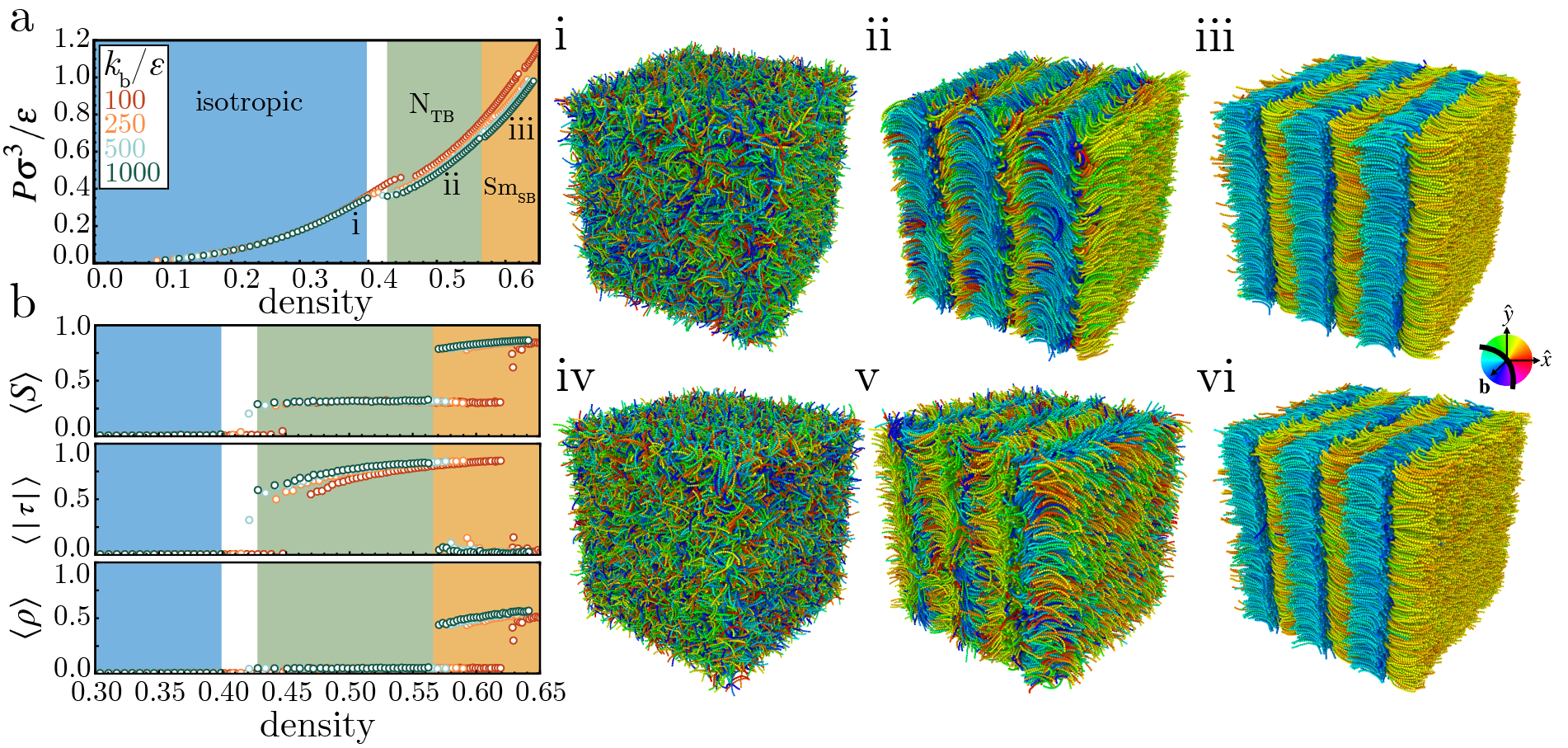}
\caption{Effect of bend elastic constant on phase diagram of colloidal bent core liquid crystals. (\textbf{a}) Pressure versus density curve for several different values of the bend elastic constant $k_{\rm b}$. (\textbf{b}) Nematic, helical and smectic order parameters plotted as function of density for several different values of $k_{\rm b}$. The colored regions in (\textbf{a}) and (\textbf{b}) drawn for the case of the most rigid rods with $k_{\rm b}=1000\epsilon$. Simulation snapshots of the isotropic, nematic twist-bend and smectic splay-bend phases provided for stiff (i.e. $k_{\rm b}=1000\epsilon$) rods in i-iii and flexible rods ($k_{\rm b}=100\epsilon$) in iv-vi for $N=40500$ rods, $k_{\rm B}T/\epsilon=1$, $P\sigma^3/\epsilon=0.3,0.5,0.8$ for i-iii and $P\sigma^3/\epsilon=0.3,0.8,1.1$ for iv-vi. Rods are colored by the direction the $xy$-projection of their bend director, $\mathbf{b}$, points to when centered on the color wheel, as illustrated in the legend.} \label{fig: phases}
\end{figure*}

Throughout this article, simulations are initialized with a high density ordered state (see Fig.~\ref{fig: Initial States}a) which is obtained by replicating a much smaller ordered phase to obtain a larger one of desired size. Specifically, the small state is created by placing $1500$ rigid rods in a cubic cell with $L = 100\sigma$, increasing the pressure to $P=1.0\epsilon/\sigma^3$ and allowing the system to equilibrate over a time of $\sim1.5\cdot10^5\tau_{\rm LJ}$, which is sufficient for the system to form a defect-free ordered state. The resulting state is used as a unit cell to create a $3\times 3$ $\times 3$ replica with $40500$ rods. Lastly, this new state is allowed to relax via a short isobaric simulation using the bonded particle model. Using this, ordered phases with other densities can be obtained by changing the pressure and allowing the system to equilibrate. An example of such an ordered state, with $P=0.5\epsilon/\sigma^3$, is given in Fig.~\ref{fig: Initial States}b. Simulations are initialized in this way\textemdash rather than randomly placing $40500$ rods in the simulation cell (Fig.~\ref{fig: Initial States}c)\textemdash to avoid getting stuck in disordered states characterized by networks of Skyrmion-like defect lines. An example of a disordered state, prepared in this way, is given in Fig.~\ref{fig: Initial States}d. These so called amorphous blue phases have previously been reported in MD simulations of rigid bent-core liquid crystals~\cite{subert2024achiral} and are at-least strongly metastable in bulk. This is evidenced by the observation that, while these Skyrmion-like defects can be annealed away for small systems with $1500$ rods, they remain in the larger system over the time scales that are currently accessible. Thus, as the initially isotropic simulations are subject to prohibitively long relaxation times for all but the smallest systems, the high density ordered phase is used as the initial state for the study presented here. While the initialization scheme outlined here effectively avoids this meta-stable trap, it is worth noting that it comes with the tradeoff that the final states inherit the average global orientation of the initial state.


\section{Results}\label{sec: results}

In this work, we are primarily interested in studying the effect that elastic variations in curvature have on the ordered phases of colloidal bent-core liquid crystals. As the bonded particle model has a large  parameter space, we focus on fixed, large values of $k_{\rm r}=1000\epsilon/\sigma^2$, $k_{\rm s}=500\epsilon/\sigma^2$ and $k_{\rm t}=1000\epsilon$ and vary the bending modulus, $k_{\rm b}/\epsilon=[100,250,500,1000]$. This range of parameters corresponds to the case of bendable rods that are inextensible in arc-length and very stiff with respect to shear and twist.  Here it is relevant to point out that the bending stiffness can't be made arbitrarily small compared to the other deformation modes, as the BPM becomes unstable for large ratios of the elastic constants. Using these parameters, we equilibrate systems with $N=40500$ rods at temperature $T=\epsilon/k_{\rm B}$ and pressures ranging from $P=0.1$ to $1.1\epsilon/\sigma^3$. We then measure the number density $\phi$ as a function of pressure, which is varied in increments of $\Delta P=0.1\epsilon/\sigma^3$. The resulting pressure versus density curves are plotted in Fig.~\ref{fig: phases}a. The phase of each simulation is determined by measuring the average nematic $\langle S\rangle$, helical $\langle \vert\tau\vert\rangle$ and smectic $\langle\rho\rangle$ order parameters. In this notation, $\langle\cdots\rangle$ denotes an ensemble average taken over equilibrium configurations. The nematic order parameter is defined as the largest eigenvalue of the average $\mathbb{Q}$-tensor~\cite{de1993physics}
\begin{equation}
\mathbb{Q}=\frac{1}{N}\sum_{i}^N\bigg(\frac{3}{2}\mathbf{n}_i\otimes\mathbf{n}_i-\frac{1}{2}\mathbb{I}\bigg)
\end{equation}
where $\otimes$ denotes the outer product of two vectors and $\mathbb{I}$ is the identity matrix. Similarly, the helical order parameter $\tau$ is defined as the largest eigenvalue of the tensor~\cite{xu2020general,subert2024achiral}
\begin{equation}
\mathbb{M}=\frac{1}{N}\sum_{i}(\mathbf{n_i}\otimes\mathbf{m_i}+\mathbf{m_i}\otimes\mathbf{n_i})
\end{equation}
where $\mathbf{n}_i$ is the nematic director and $\mathbf{m}_i$ is the out-of-plane director of the $i$th rod. Lastly, the smectic order parameter is defined as~\cite{de1993physics}
\begin{equation}
\langle \rho\rangle=\frac{1}{N}\sum_{j}^N e^{\frac{2\pi i}{d}\mathbf{n}_{\rm avg}\cdot\mathbf{r}_j}
\end{equation}
where $d$ is the optimal layer spacing that maximizes $\langle \rho\rangle$, $\mathbf{n}_{\rm avg}$ is the average nematic director and $\mathbf{r}_j$ is the position of the mid-point along the backbone of the $j$th rod. Here, we take $\mathbf{n}_{\rm avg}$ to be the eigenvector associated with the largest eigenvalue of the $\mathbb{Q}$-tensor. 

The three order parameters are plotted as a function of density in Fig.~\ref{fig: phases}b. For the stiffest rods (i.e. $k_{\rm b}=1000\epsilon$), at low density and pressure, the rods have negligible amounts of order, corresponding to an isotropic gas of curved rods (see Fig.~\ref{fig: phases}i). At intermediate density and pressure, there is a large amount of helical order, moderate amounts of nematic order and very little smectic order. This is consistent with the appearance of a nematic twist-bend phase (see Fig~\ref{fig: phases}ii). Additionally, the isotropic and $N_{\rm TB}$ phases are separated by a discontinuous jump in density, indicative of a first-order transition. At high density and pressure, there is vanishing helical order, a large amount of nematic order and intermediate amounts of smectic order. This corresponds to the appearance of a smectic splay-bend phase (see Fig~\ref{fig: phases}iii). It is important to note that this observed sequence of phases, under increasing pressure and density, agrees with previous studies of rigid bent-core liquid crystals that have a similar opening angle.\cite{subert2024achiral} However, the overall equation of state observed here is shifted to lower density compared to this previous study. This shift can be accounted for by noting that the earlier studies of rigid rods also used a much stiffer, nearly hard-sphere potential in comparison to the softer Lennard-Jones potential used here.

\begin{figure}[ht!]%
\includegraphics[width=0.5\textwidth]{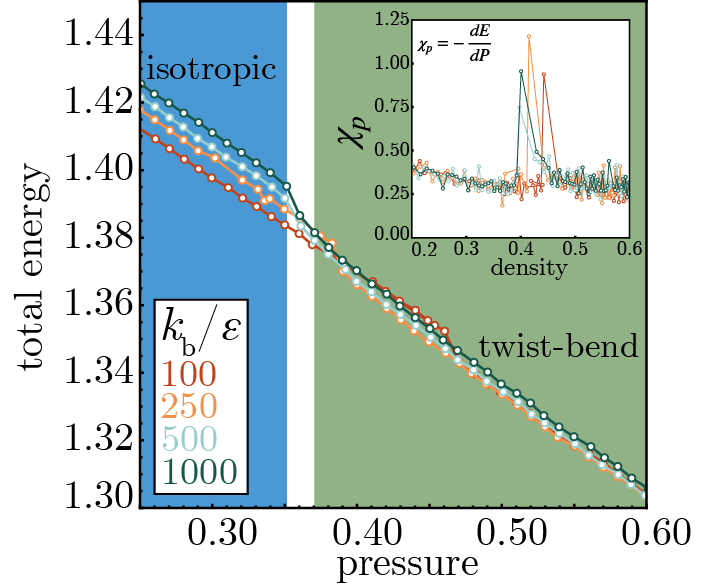}
\caption{Total energy as function of pressure for several different values of elastic constant $k_{\rm b}$. Colored background denotes the phases for rigid ($k_{\rm b}=1000\epsilon$) rods. Inset shows the susceptibility across the isotropic to nematic twist-bend transition.}  \label{fig: total energy}
\end{figure}

Upon decreasing the bend elastic constant $k_{\rm b}$, the same sequence of ordered phases is observed as shown in Fig.~\ref{fig: phases}iv-vi for  the isotropic, twist-bend and splay-bend phases with $k_{\rm b}=100\epsilon$. However, both the transition to the nematic twist-bend and smectic splay-bend phase are increasingly shifted to higher density as $k_{\rm b}$ decreases. In addition, an apparent decrease is observed in the density gap separating the $I$-$N_{\rm TB}$ phase as the bending flexibility (i.e. by lowering $k_{\rm b}$) increases. This effect can be seen more clearly in the total energy as a function of pressure, as shown in Figure~\ref{fig: total energy}, which suggests that the bend elasticity weakens the first-order transition between the isotropic and nematic twist-bend phase.
\subsection{Fluctuations in Rod Curvature}\label{sec: fluctuations}
\begin{figure}[ht!]%
\includegraphics[width=0.5\textwidth]{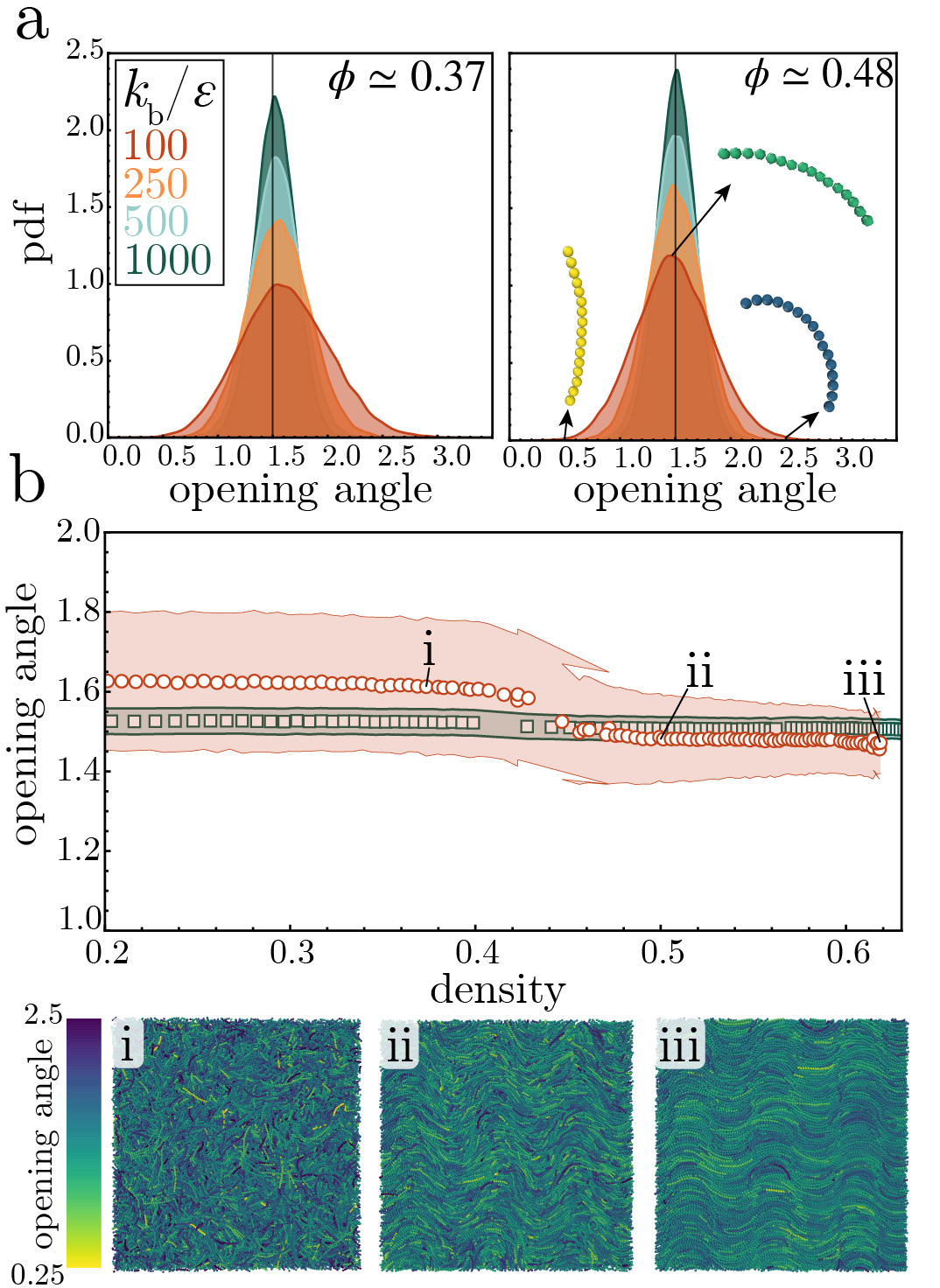}
\caption{Effect of bending constant and density on opening angle of bent colloidal rods. ($\mathbf{a}$) Distribution of rod opening angles for four values of the bending constant. The left histogram corresponds to a low density ($\phi\simeq0.37\sigma^{-3}$) isotropic state and the right histogram corresponds to an intermediate density ($\phi\simeq0.48\sigma^{-3}$) nematic twist bend phase. The vertical black line denotes the prescribed opening angle $\Psi=1.5$. Example of simulated rods with different opening angles given in the inset of the right histogram. ($\mathbf{b}$) Mean opening angle of the $k_{\rm b}=100\epsilon$ (red circles) and $k_{\rm}=1000\epsilon$ (green squares) as a function of density. Colored region around the mean denotes the standard deviation in opening angle. Example states of each of the main phases, with rods colored by opening angle shown in $\mathbf{i}$-$\mathbf{iii}$. }  \label{fig: bend fluctuations}
\end{figure}

To understand the role that bend elasticity plays in driving the isotropic to nematic twist-bend and nematic twist-bend to smectic splay-bend transitions to higher density, we examined curvature fluctuations as a function of $k_{\rm b}$. Here, the radius of curvature of a rod is determined by measuring the backbone length, $L$, and the chordal length, $C$, between the ends of the rods and solving the transcendental equation
\begin{equation}
C=2R\sin\bigg(\frac{L}{2R}\bigg)
\end{equation}
for the radius of curvature $R$. In Fig.~\ref{fig: bend fluctuations}a the distribution of opening angle $\Psi=L/R$ is shown for several values of the density and bending constant. These results show a strong increase in curvature fluctuations as the  bending modulus is decreased, with the mode of the distribution remaining largely fixed at the target value of $\Psi=1.5$. In Fig.~\ref{fig: bend fluctuations}b, the average opening angle as a function of density is shown. The fluctuations in bend are larger in the low density isotropic phase, with the distribution skewing slightly towards larger $\Psi$ (i.e. more curvature). Furthermore, as the density increases, the mean and standard deviation of the opening angle decrease across the transition to the nematic twist-bend phase with the mean becoming slightly less than the target. This suggests that fluctuations in bend are effectively suppressed at high density, where there is less room for the rods to vary their shape and the rods are forced to flatten out slightly. Additionally, this difference in the mean opening angle decreases with increasing stiffness, with the $k_{\rm b}=1000\epsilon$ rods exhibiting little to no change in average curvature over the entire density range considered, indicating that they are effectively rigid. Examples of each phase, with individual rods colored by their opening angle, are given in Fig.~\ref{fig: bend fluctuations}i-iii for $k_{\rm b}=100\epsilon$. Here, no coherent gradients in opening angle are observed across all three phases, with the fluctuations apparently homogeneous in space.

\subsection{Structural Effects of Shape Elasticity}\label{sec: structure}
\begin{figure*}[ht!]%
\includegraphics[width=\textwidth]{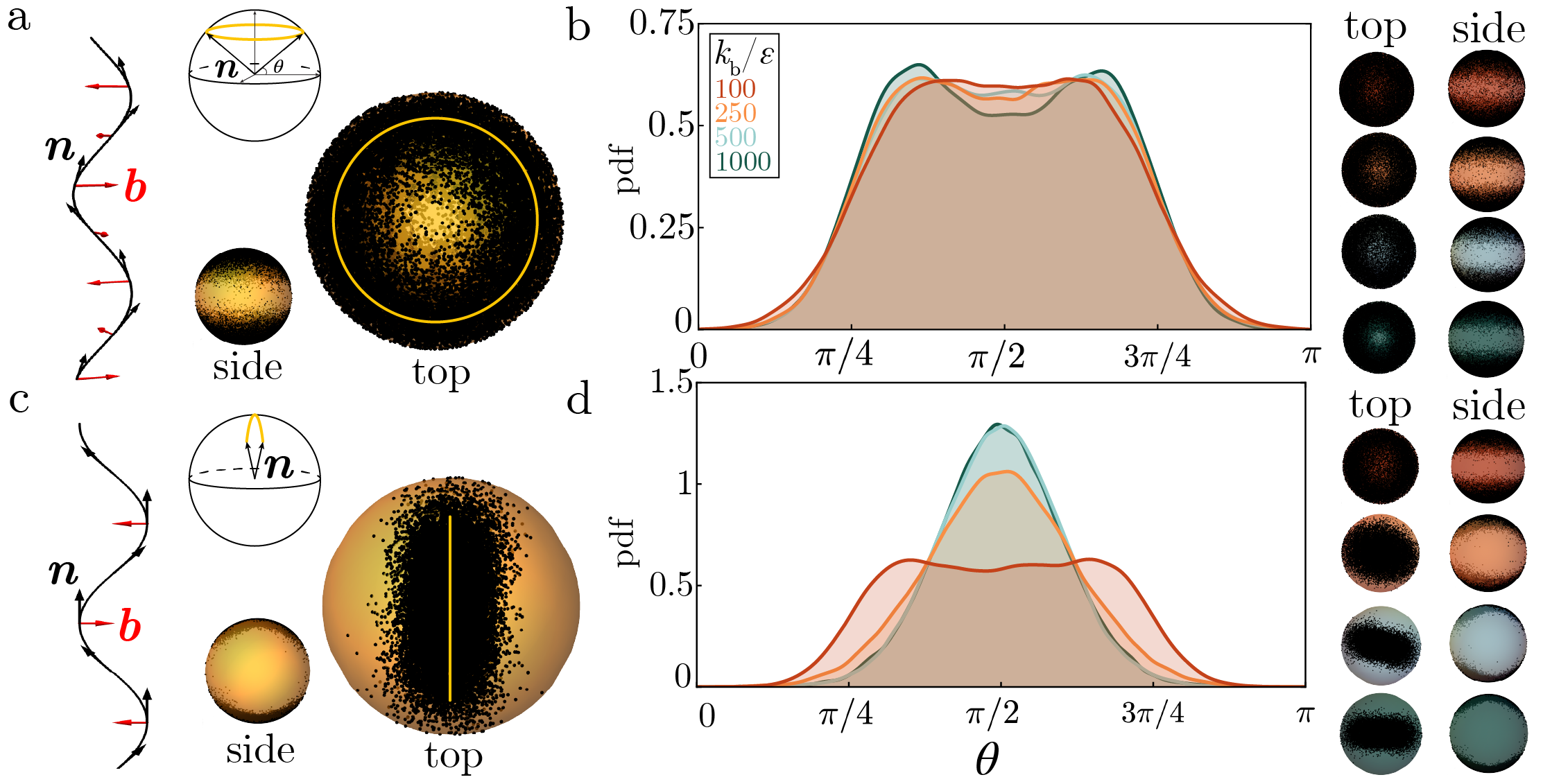}
\caption{ Flexibility induced fluctuations around ground state ordered phases. (\textbf{a}) Heliconical integral curves for nematic twist-bend phase with nematic director illustrated as black arrows. Bend director also drawn (red arrows) to highlight three-dimensional nature of the twist-bend helices. Gauss mapping of nematic director drawn as gold ring on unit sphere. Side and top view of Gauss mapping of rod directors taken from simulation in nematic twist-bend phase (with $k_{\rm b}=1000\epsilon$) shown for comparison. Here, each black point on the sphere corresponds to a single rod \textit{in silico}. (\textbf{b}) Angular distribution of rod orientation shown for several different values of elastic constant, $k_{\rm b}$. Simulations run at $P\sigma^3/\epsilon=0.5$ to target twist-bend phase. Angle $\theta$ defined as the angle above the equatorial plane that rod director points to when mapped to unit sphere. Side and top views of Gauss mapping of rods provided for each value of $k_{\rm b}$. (\textbf{c}) Integral curve for splay-bend phase with nematic director illustrated as black arrow. Bend director also illustrated (red arrows). Here, the integral curve lies within two-dimensional plane. Gauss mapping of nematic director illustrated as gold line segment on unit sphere. Side and top view of Gauss mapping of rod directors taken from simulation in smectic splay-bend phase (with $k_{\rm b}=1000\epsilon$ and $P\sigma^3/\epsilon=0.8$) shown for comparison. (\textbf{d}) Angular distribution of rod orientation shown for several different values of elastic constant, $k_{\rm b}$. Simulations run at $P\sigma^3/\epsilon=0.8$ so that system would, at least, be in splay-bend phase for stiff rods (i.e. $k_{\rm b}=1000\epsilon$). The angle $\theta$ is defined as the angle above the equatorial plane that rod director points to when mapped to unit sphere. Side and top views of Gauss mapping of rods provided for each value of $k_{\rm b}$. Here the color of the sphere denotes the specific value of the $k_{\rm b}$ and follows the same color mapping scheme denoted in the legend of subplot (\textbf{b}).} \label{fig: structural fluctuations}
\end{figure*}

In this section, the effects that flexibility induced fluctuations in rod curvature have on the ordered phases of bent-core liquid crystals are investigated. These effects are quantified by considering the Gaussian map of the rod directors, $\mathbf{n}$, onto the unit sphere and comparing it to an idealized twist- or splay-bend director field. This allows us to directly quantify how much the random variations in curvature cause each phase to deviate from the idealized structure. For example, the nematic twist-bend phase has director field~\cite{jakli2018physics,dozov2001spontaneous}
\begin{equation}
\mathbf{n}=\cos\theta\mathbf{e}_z+\sin\theta\big[\cos qz\mathbf{e}_x+\sin qz \mathbf{e}_y\big]
\end{equation}
yielding a heliconical integral curve, with wave-vector $q$, and tracing out an annular halo at polar angle $\theta$ on the unit sphere. A schematic illustration of this integral curve and its Gauss map is given in Fig.~\ref{fig: structural fluctuations}a. Comparing this to the Gauss map for a simulation in the nematic twist-bend phase, we clearly see the expected halo pattern, however, the annuli has some width corresponding to fluctuations around the ideal structure. This effect can be analyzed by looking at the distribution of the polar angle (defined with respect to the equatorial plane of the unit sphere) that the rods point to on the unit sphere (see Fig.~\ref{fig: structural fluctuations}b). From here, we see a clear double peaked structure in the stiff (i.e. $k_{\rm b}=1000\epsilon$) rod's distribution. As the elastic constant is decreased, the height of the peaks relative to the valley between them decreases and they eventually disappear, leaving a more uniform distribution of rod orientation. This suggests that, at fixed density, flexibility induced curvature fluctuations de-stabilize the nematic twist-bend phase and drive the system into the isotropic phase at low enough $k_{\rm b}$.

A similar analysis is done for the splay-bend phase, which has director field~\cite{jakli2018physics,chaturvedi2019mechanisms}:
\begin{equation}
\mathbf{n}=\sin(\theta\sin qz)\mathbf{e}_x+\cos(\theta\sin qz)\mathbf{e}_z
\end{equation}
which has integral curves that form planar arcs that oscillate through angle $\theta$ with wave number $q$. Here, the nematic director traces out a finite arc with length $\theta$ when mapped onto the unit sphere. Again, this is compared to the Gauss map of the simulated rod directors (see Fig.~\ref{fig: structural fluctuations}c). Here, it is apparent that the mapping for the $k_{\rm b}=1000\epsilon$ rods yields a finite length and width ribbon pattern, consistent with the simulation being in the splay-bend phase. Looking at the distribution of rod polar angles (see Fig.~\ref{fig: structural fluctuations}d), we see a clear single peaked structure, indicating that the rod orientation is Gaussian distributed around the ideal splay-bend phase. As $k_{\rm b}$ is lowered, the width of this distribution initially widens, while maintaining its Gaussian form. However, as the rods become more flexible, the distribution splits into a double peaked structure like that observed for the twist-bend phase. Thus, it is observed that the splay-bend phase also becomes unstable under increased curvature fluctuations and eventually transitions to the nematic twist-bend phase as the bending moduli is decreased.

\section{Discussion and Conclusion}\label{sec: conclusion}

In this article, a bonded particle model is used to describe the shape elasticity of curved rods. This allowed us to perform molecular dynamics simulations to investigate the effect of bend deformations on the equilibrium phase behavior of colloidal bent-core liquid crystals. Specifically, we focused on the intermediate curvature regime, where the rods have an undeformed opening angle of $\Psi=1.5$. Further, we considered a specific cut through the bonded particle model parameter space intended to mimic inextensible rods that are flexible in curvature but stiff to out-of-plane twisting and shear. Varying the pressure, we observe a low density isotropic phase that goes through a first order transition to a nematic twist-bend phase at intermediate density. Then, at even higher density, the system transitions to a smectic splay-bend phase. This same sequence of phases has been previously reported in studies of rigid rods with a similar opening angle.\cite{kubala2022silico,chiappini2021generalized,subert2024achiral}

In comparison to the case of rigid rods, less smectic ordering in both the twist- and splay-bend phases is observed,\cite{chiappini2021generalized} suggesting that rod flexibility  plays a similar role as dispersity in reducing periodic density modulation.\cite{chiappini2019biaxial} We also found that lowering the rod's bending stiffness $k_{\rm b}$ shifted both of the observed transitions to higher density and weakened the first order isotropic-nematic twist-bend transition. To explain this, we examined the effect of bending stiffness $k_{\rm b}$ and density $\phi$ on the distribution of rod curvatures and found that increased flexibility (i.e. lower $k_{\rm b}$) leads to a wider variation in curvature and that these effects are more pronounced in the isotropic phase. At higher density, the curvature fluctuations are suppressed, and the mean opening angle is lower than the preferred value, suggesting that the rods are slightly flattened from their undeformed shape. Lastly, we show that, at fixed density, these curvature fluctuations lead to increased deviation from the ideal director field, resulting in a destabilization of the ordered phases. Taken with the observation that fluctuations are suppressed by increasing density, this effect accounts for the shift in phase boundary to higher density with decreasing elastic constants.

While this work represents a first investigation into the effects of shape elasticity on the behavior of curved colloidal rods, much work remains. For example, the results presented here correspond to a specific cut through the elastic parameter space of the bonded particle model. Although this choice was made to isolate the effects of bending from other deformation modes that allow the rods to escape frustration by changing their shape entirely, it is far from the only interesting choice one could make. For example, one could equally fix the stretching parameter and uniformly vary the shearing, twisting, and bending constants. This would be akin to altering properties of an inextensible rod. In fact, the parameters of this bonded particle model can be mapped onto the material parameters of an elastic rod~\cite{chen2022comparative} (i.e. the Young's modulus, Poisson ratio, rod length and rod diameter), allowing the behavior of curved rods with specific material properties to be studied.

Similarly, the work presented here corresponds to a specific, fixed curvature cut through the density versus opening angle plane of phase space. Decreasing the bending stiffness of rods with lower opening angles (i.e. lower curvature) might cause the rods to flatten out and behave as if they were straight rods. In contrast, in the limit of larger opening angle, flexibility could encourage the rods to close into rings that pack into columnar stacks.

Finally, we point out that geometric frustration has been theorized to induce structural gradients of bend in both bent-core liquid crystals~\cite{niv2018geometric,meiri2021cumulative,hackney2023dispersed} and similar systems of curved colloidal shells~\cite{tanjeem2022focusing,sullivan2024self}. However, we observe that the fluctuations in rod curvature are randomly distributed in space and don't form coherent gradients. This discrepancy is likely the result of specific rod-rod interaction used throughout this work. For example, in systems of cohesive particles,  it has been shown that the effects of frustration are dealt with via a combination of inter-particle interaction compliance and intra-particle shape deformation.\cite{sullivan2024self} In this situation, the propagation of frustration induced strain gradients is known to be controlled by the ratio of the cohesive strength to the shape elastic stiffness. Furthermore, the presence of these structural gradients in geometrically frustrated assembly~\cite{hagan2021equilibrium} give rise to a rich phase space characterized by polydisperse states of self-limited assembly at low density, bulk lattices of topological defects at high density and an intermediate density gel phase.\cite{bj18-bphb} This situation could also provide a mechanism for stabilizing liquid crystalline blue phases, as the constituent double twist or Skyrmion-like defects require gradients of bend.\cite{binysh2020geometry} Overall, the bonded particle model employed throughout this work provides an excellent framework for future numerical studies of the effects of shape elasticity in colloidal bent core liquid crystals as well as other soft matter systems.

\section*{Conflicts of interest}
There are no conflicts to declare.

\section*{Data availability}
The data from the molecular dynamics simulations and associated analysis are available from the authors upon reasonable request.

\section*{Acknowledgements}

The authors thank Thomas O'Connor and Ting Ge for many helpful discussions. This work was performed in part at the Center for Integrated Nanotechnologies, an Office of Science User Facility operated for the U.S. Department of Energy (DOE) Office of Science. Sandia National Laboratories is a multimission laboratory managed and operated by National Technology \& Engineering Solutions of Sandia, LLC, a wholly owned subsidiary of Honeywell International, Inc., for the U.S. DOE’s National Nuclear Security Administration under contract DE-NA-0003525. The views expressed in the article do not necessarily represent the views of the U.S. DOE or the United States Government.

\bibliography{apssamp}

\end{document}